# Bayesian model reduction

*Karl Friston, Thomas Parr and Peter Zeidman*

*The Wellcome Centre for Human Neuroimaging, Institute of Neurology, University College London, UK.*

**Correspondence**: *Peter Zeidman*
*The Wellcome Centre for Human Neuroimaging*
*Institute of Neurology*
*12 Queen Square, London, UK WC1N 3AR*
*peter.zeidman@ucl.ac.uk*

**Summary**

This paper reviews recent developments in statistical structure learning; namely, Bayesian model reduction. Bayesian model reduction is a method for rapidly computing the evidence and parameters of probabilistic models that differ only in their priors. In the setting of variational Bayes this has an analytical solution, which finesses the problem of scoring large model spaces in model comparison or structure learning. In this technical note, we review Bayesian model reduction and provide the relevant equations for several discrete and continuous probability distributions. We provide worked examples in the context of multivariate linear regression, Gaussian mixture models and dynamical systems (dynamic causal modelling). These examples are accompanied by the Matlab scripts necessary to reproduce the results. Finally, we briefly review recent applications in the fields of neuroimaging and neuroscience. Specifically, we consider structure learning and hierarchical or empirical Bayes that can be regarded as a metaphor for neurobiological processes like abductive reasoning.

**Keywords**: Bayesian methods, empirical Bayes, model comparison, neurobiology, structure learning

# 1. Introduction

Over the past years, Bayesian model comparison and structure learning have become key issues in the neurosciences (Collins and Frank, 2013, Friston et al., 2017a, Salakhutdinov et al., 2013, Tenenbaum et al., 2011, Tervo et al., 2016, Zorzi et al., 2013); particularly, in optimizing models of neuroimaging time series (Friston et al., 2015, Woolrich et al., 2009) and as a fundamental problem that our brains have to solve (Friston et al., 2017a, Schmidhuber, 1991, Schmidhuber, 2010). Indeed, it is often said that the answer to every question is Bayesian model comparison. This notion has a deep truth to it; in the sense that any question – that can be framed in terms of competing hypotheses – can only be answered by appeal to the evidence for those hypotheses. In other words, the answer to any question reduces to a comparison of hypotheses or model evidence, implicit in the use of Bayes factors, or differences in log evidence (Kass and Raftery, 1995). This technical note reviews a relatively new procedure called Bayesian model reduction that is now used routinely in neuroscience for both data analysis (Friston and Penny, 2011, Friston et al., 2015, Friston et al., 2016) and theoretical neurobiology (Friston et al., 2017a). The basic idea behind Bayesian model reduction is straightforward and may find useful applications in a variety of fields.

Consider a statistical model of some data $y$ with parameters $\theta$. A generative or forward model of the data comprises two parts – the likelihood of observing some data under a particular setting of the parameters $P(y|\theta)$ and a prior probability density $P(\theta)$. The balance between the two determines the degree to which a parameter is free to explain data. For instance, a parameter that has a Gaussian prior with high variance (low precision) will be readily informed by the data, whereas a parameter with low prior variance (high precision) will be constrained to values near its prior expectation. This use of priors, to effectively 'switch off' parameters – by fixing them at some pre-determined value (e.g. zero) – will be important in what follows. After specifying priors, Bayesian inference – also called estimation or inversion – furnishes a posterior probability over the parameters $P(\theta|y)$. These posteriors are related to the priors and likelihood by Bayes rule:

$$P(\theta|y) = \frac{P(y|\theta)P(\theta)}{P(y)}$$
$$P(y) = \int P(y|\theta)P(\theta)d\theta \quad (1)$$

To compare different models of the same data – i.e., to perform *Bayesian model comparison* – it is necessary to evaluate the evidence for each model $P(y)$. This is the probability of sampling some data under a particular model, also called the integrated or marginal likelihood.

Generally speaking, evaluating this quantity is an intractable analytic problem (due to the integral on the second line of Eq. 1), which is typically resolved using approximate Bayesian inference. These methods have non-trivial evaluation times on typical computer hardware. Indeed, even the most efficient *variational* approaches, detailed in the next section, operate on the order of seconds for simple models with small datasets and hours or days for the most challenging problems. This computational cost can preclude the evaluation of large model spaces in reasonable time. This problem is addressed in the setting of approximate Bayesian inference by the introduction of an efficient scheme for analytically scoring the evidence for large numbers of models.

The situation we are concerned with here is as follows. Imagine that we have used some Bayesian inference scheme to estimate the parameters of a model; i.e. we have optimized a posterior probability density over the parameters, given prior beliefs and some data. We now want to consider alternative models defined in terms of alternative prior beliefs. Usually, these would constitute reduced models with more precise or informative priors, which constrain or completely eliminate some (mixture of) free parameters. Bayesian model reduction provides a way of evaluating the evidence of the reduced model based on, and only on, the original (parent or full) priors and approximate posteriors. In other words, it allows one to evaluate the evidence for a new set of priors based upon the original estimates. This follows the same sort of procedure used in classical inference, where *F* tests are based on, and only on, the parameters of a general linear model. An obvious application is scoring large model spaces after inverting a parent model with relatively uninformative (i.e., flat) priors. One can now revisit model assumptions by scoring alternative models in which various combinations of parameters are suppressed or eliminated using precise (i.e., reduced) priors. This enables one to remove redundant parameters and prune models, via the exploration of large model spaces: in other words, it enables an efficient form of structure learning.

In what follows, we review the tenets of variational Bayes, introduce the notion of reduced free energy and review particular forms for continuous and discrete models. The subsequent sections provide worked examples of Bayesian model reduction in the context of linear regression, Gaussian mixture models and dynamic models typically used for modelling timeseries data. Finally, some empirical applications are reviewed from the field of neuroimaging and neurobiology (that interested readers can pursue in the original publications).

## 2. Variational Bayes

The log of the model evidence $\ln P(y)$, hereafter *log evidence*, scores the quality of a model and is the basis for Bayesian model comparison. However, as set out in the introduction, the log evidence and the ensuing posteriors cannot, in general, be computed analytically. In many instances, this is resolved by converting the difficult integration problem inherent in Eq. 1 into a tractable optimization problem; namely, maximising a variational (free energy) bound on log evidence (Beal, 2003, Fox and Roberts, 2012, Kschischang et al., 2001, MacKay, 1995, Yedidia et al., 2005). This variational free energy is known in machine learning as an evidence lower bound (ELBO). The associated optimization is known variously as approximate Bayesian inference, variational Bayes and variational ensemble learning, with many special cases such as exact Bayesian inference, variational Laplace, Bayesian (e.g., Kalman) filtering, expectation maximisation, variational message passing, belief propagation, and so on (Dauwels, 2007, Kschischang et al., 2001, Loeliger, 2002, Mitter and Newton, 2003, Roweis and Ghahramani, 1999, Yedidia et al., 2005). In brief, nearly every (variational) approach to Bayesian model inversion and comparison can be expressed as optimizing a variational free energy functional of some data and an approximate posterior distribution or density (denoted herein as $Q$). The question we focus on is how this variational free energy can be computed quickly and efficiently, under a change of priors – or when adding hierarchical constraints to evaluate a deep or hierarchical model of some data.

To briefly reprise, variational Bayes involves identifying a probability density $Q(\theta)$ that approximates the posterior $P(\theta | y)$. An optimisation is performed to identify a density that minimizes the difference between the approximate and true posterior. The difference measure typically used is the Kullback-Leibler divergence:

$$D_{KL}\left[Q(\theta) \| P(\theta | y)\right] = E_Q\left[\ln Q(\theta) - \ln P(y, \theta)\right] + \ln p(y) \qquad (2)$$

This cannot be computed directly, as it depends on the log evidence $\ln p(y)$, which contains the difficult integral in Eq. 1. However, one can compute a related functional, called the variational free energy $F$ (hereafter the free energy), and use this to score the quality of the approximation:

$$F[P(\theta)] = E_Q[\underbrace{\ln P(y, \theta)}_{\text{Energy}} - \underbrace{\ln Q(\theta)}_{\text{Entropy}}] \qquad (3)$$

This is the same as the Kullback-Leibler divergence up to a constant; i.e. the log evidence, which is constant with respect to $Q$. The name *free energy* derives from statistical physics, and by analogy the two terms in Eq. 3 are referred to as the *energy* and *entropy*. These are the expected log likelihood of some data and parameters – under an approximate posterior – and the entropy of that posterior.

By expressing the free energy as a functional of some prior beliefs, Eq. 3 emphasises that for any given model or prior belief, a free energy functional of the (approximate) posterior is well defined. By re-arrangement, some useful properties of the free energy are made apparent:

$$F[P(\theta)] = \underbrace{\ln P(y)}_{\text{Log evidence}} - \underbrace{D_{KL}[Q(\theta) \| P(\theta | y)]}_{\text{Bound}}$$
$$= \underbrace{E_Q[\ln P(y | \theta)]}_{\text{Accuracy}} - \underbrace{D_{KL}[Q(\theta) \| P(\theta)]}_{\text{Complexity}}$$
(4)

The first line shows that free energy is the log evidence minus the Kullback-Leibler divergence (i.e., nonnegative bound) between the approximate and true posterior. This means that maximising the free energy makes the approximate posterior as close as possible to the true posterior. If the assumed form of the approximate posterior coincides with the true posterior, then we have exact Bayesian inference and the free energy becomes the log evidence. The second line shows that free energy can also be expressed as *accuracy* minus *complexity*, where complexity is the divergence between the approximate posterior and prior beliefs.

Typically, the approximate posterior is factorised into marginals over subsets of the unknown parameters: $Q(\theta) = Q(\theta_1)Q(\theta_2)\ldots$. For example, parameters governing the mean and variance of some data may be assumed to be independent of one another. This is known as a mean field approximation and can greatly simplify model inversion; namely, the maximisation of free energy with respect to the marginal posteriors. Using variational calculus, it is straightforward to show that the approximate posterior of any parameter subset is the expected log probability under its Markov blanket:

$$\delta_{Q_i} F[P(\theta)] = 0$$
$$\Rightarrow \ln Q(\theta_i) = E_{Q_{\setminus i}}[\ln P(y, \theta)] \approx \ln P(\theta_i | y)$$
$$\Rightarrow Q(\theta_i) = \sigma(E_{Q_{\setminus i}}[\ln P(y, \theta)]) \approx P(\theta_i | y)$$
$$\Rightarrow F[P(\theta)] \approx \ln P(y)$$
(5)

Where $\delta_{Q_i} F[P(\theta)]$ is the functional derivative of the free energy under the *i*-th subset of parameters, $\sigma$ denotes a softmax function or normalised exponential and $Q_{\setminus i}$ denotes the Markov blanket of the *i*-th subset (Beal, 2003). This says that when the free energy is maximised, the best approximation of the posterior for each partition $i$ is the expectation of the log joint probability of the data and parameters. This expectation is taken under the parameters of all other partitions $Q_{\setminus i}$ which are held fixed. Thus, by maximising the free energy using any suitable optimisation scheme; e.g. (Friston et al., 2007, Ranganath et al., 2014), $Q(\theta_i)$ and $F$ become approximately equal to the posterior and log evidence respectively. Note that for what follows, the only prerequisites are to have estimated the posteriors and the free energy of a model – optimisation schemes that do not apply a mean field approximation are equally suitable (c.f. Ranganath et al., 2016, Rezende and Mohamed, 2015, Salimans et al., 2015).

## 3. Bayesian model reduction and reduced free energy

We now turn to the problem addressed by Bayesian model reduction: evaluating the free energy and parameters of a reduced model, based on the priors and posteriors of a full model. Consider Bayes rule expressed for a reduced model, in which the original priors $P(\theta)$ have been replaced with reduced priors $\tilde{P}(\theta)$:

$$\tilde{P}(\theta \mid y) = \frac{P(y \mid \theta)\tilde{P}(\theta)}{\tilde{P}(y)}$$
$$\tilde{P}(y) = \int \tilde{P}(y \mid \theta)\tilde{P}(\theta)\,d\theta \tag{6}$$

Here the posteriors and evidence for the reduced model are $\tilde{P}(\theta \mid y)$ and $\tilde{P}(y)$ respectively. Note that the likelihood for the full and reduced models are identical, i.e. the form of these models is assumed to be the same. By re-arrangement, we have the dual application of Bayes rule to the full and reduced models:

$$\frac{\tilde{P}(\theta \mid y)\tilde{P}(y)}{P(\theta \mid y)P(y)} = \frac{P(y \mid \theta)\tilde{P}(\theta)}{P(y \mid \theta)P(\theta)} \Rightarrow \tilde{P}(\theta \mid y) = P(\theta \mid y)\frac{\tilde{P}(\theta)P(y)}{P(\theta)\tilde{P}(y)} \tag{7}$$

This is Bayes rule expressed with reduced and original priors. We have then re-expressed these as a posterior odds ratio, so that the likelihoods cancel. The last expression shows that the posteriors of the reduced model can be expressed in terms of the posteriors of the full model and the ratios of priors and evidences. By integrating over the parameters in Eq. 7, we get the evidence ratio of the reduced and full models (where the left hand side of the reduced posterior integrates to unity):

$$\int \tilde{P}(\theta \mid y)\,d\theta = 1 = \int P(\theta \mid y)\frac{\tilde{P}(\theta)P(y)}{P(\theta)\tilde{P}(y)}\,d\theta$$
$$\Rightarrow \tilde{P}(y) = P(y)\int P(\theta \mid y)\frac{\tilde{P}(\theta)}{P(\theta)}\,d\theta \tag{8}$$

Taking the log of Eq. 8 and substituting approximate values for the model posterior and log evidence introduced in Eq. 5, we get the log evidence for the reduced model:

$$\ln \tilde{P}(y) = \ln \int Q(\theta) \frac{\tilde{P}(\theta)}{P(\theta)} d\theta + \ln P(y)$$

$$\Rightarrow \qquad (9)$$

$$F[\tilde{P}(\theta) : P(\theta)] \approx \ln E_Q\left[\frac{\tilde{P}(\theta)}{P(\theta)}\right] + F[P(\theta)] \approx \ln \tilde{P}(y)$$

Where the notation $F[\tilde{P}(\theta) : P(\theta)]$ indicates the free energy of the reduced model derived from the full model, i.e. $F[\tilde{P}(\theta) : P(\theta)] := F[\tilde{P}(\theta)]$. We can similarly derive the approximate posterior of the parameters in the reduced model:

$$\ln \tilde{Q}(\theta) = \ln Q(\theta) + \ln \frac{\tilde{P}(\theta)}{P(\theta)} - \ln E_Q\left[\frac{\tilde{P}(\theta)}{P(\theta)}\right] \qquad (10)$$

This is obtained by substituting the free energies of the full and reduced models (Eq. 4 and Eq. 9) in place of the respective log evidences in Eq. 7. These (approximate) equalities mean one can evaluate the posterior and evidence of any reduced model, given the posteriors of the full model. In other words, $F[\tilde{P}(\theta) : P(\theta)] \approx \ln \tilde{P}(y)$ allows us to skip the optimization of the reduced posterior $\tilde{Q}(\theta)$ and use the optimized posterior of the full model to compute the evidence (and posterior) of the reduced model directly. Some readers will recognise this as a generalisation of the Savage-Dickey density ratio (Savage, 1972, Verdinelli and Wasserman, 1995) to any new prior. Crucially, in the variational setting of approximate Bayesian inference, it is straightforward to evaluate the reduced posterior analytically, because it has a known form – as we will see in the examples below.

Equations 9 and 10 can be evaluated rapidly using analytic expressions for specific (exponential) forms of the approximate posterior detailed in the next section. As its name implies, Bayesian model reduction can only be used for Bayesian model comparison when all models of interest can be cast as reduced forms of a parent or full model; in other words, the full model must contain all the parameters of any model that will be considered. This means one cannot compare models that have a completely different form. However, in practice, most model comparisons tend to be framed in terms of models with and without key (sets of) parameters. In what follows, we look at Bayesian model reduction for ubiquitous forms of the approximate posterior for models of continuous and discrete data.

## 4. Variational Laplace

Variational Laplace corresponds to approximate Bayesian inference when assuming the approximate posterior $Q(\theta) = \mathcal{N}(\mu, C)$ is Gaussian. Under this Laplace assumption, the reduced forms of the approximate posterior and free energy have simple forms: see Friston and Penny (2011) for details.

$$P(\theta) = \mathcal{N}(\eta, \Sigma)$$
$$\tilde{P}(\theta) = \mathcal{N}(\tilde{\eta}, \tilde{\Sigma})$$
$$Q(\theta) = \mathcal{N}(\mu, C)$$
$$\tilde{Q}(\theta) = \mathcal{N}(\tilde{\mu}, \tilde{C})$$

(11)

$$\tilde{C}^{-1} = \tilde{P} = P + \tilde{\Pi} - \Pi$$
$$\tilde{\mu} = \tilde{C}(P\mu + \tilde{\Pi}\tilde{\eta} - \Pi\eta)$$

$$\Delta F = \tfrac{1}{2}\ln|\tilde{\Pi} P \tilde{C}\Sigma| - \tfrac{1}{2}(\mu \cdot P\mu + \tilde{\eta} \cdot \tilde{\Pi}\tilde{\eta} - \eta \cdot \Pi\eta - \tilde{\mu} \cdot \tilde{P}\tilde{\mu})$$
$$\Delta F := F[\tilde{P}(\theta) : P(\theta)] - F[P(\theta)]$$

Here $\Pi$ and $\Sigma$ are the prior precision and covariance respectively, while $P$ and $C$ are the corresponding posterior precision and covariance. The prior expectations are $\eta$ and $\tilde{\eta}$ in the full and reduced models respectively, while $\mu$ and $\tilde{\mu}$ are the posterior expectations. The last equality in Eq. 11 defines the change in free energy $\Delta F$ that corresponds to a variational log Bayes factor. This is derived by substituting Gaussian forms for the probability density functions into Eq. 9; please see Friston and Penny (2011). Note that when a parameter is removed from the model, by shrinking its prior variance to zero, the prior and posterior moments become the same – and the parameter no longer contributes to the reduced free-energy. Effectively, Eq. 11 allows us to score any reduced model or prior in terms of a reduced free energy, while directly evaluating the posterior over its parameters. The corresponding form for discrete state space models is as follows, when the posteriors and priors have a Dirichlet distribution parameterised in terms of concentration priors.

## 5. Bayesian model reduction for discrete models

In the context of discrete models, where the posterior has a categorical distribution (and conjugate Dirichlet priors specified in terms of concentration parameters), Bayesian model reduction reduces to something remarkably simple: by applying Bayes rules to full and reduced models it is straightforward to show that the change in free energy (i.e., log Bayes factor) can be expressed in terms of posterior concentration parameters $\mathbf{a}$, prior concentration parameters $a$, the prior concentration parameters that define a reduced or simpler model $\tilde{a}$ and the ensuing reduced posterior $\tilde{\mathbf{a}}$. Using $\mathrm{B}(\bullet)$ to denote the multivariate beta function, we get (Friston et al., 2017a):

$$P(\theta) = Dir(a)$$
$$\tilde{P}(\theta) = Dir(\tilde{a})$$
$$Q(\theta) = Dir(\mathbf{a})$$
$$\tilde{Q}(\theta) = Dir(\tilde{\mathbf{a}})$$

$$\tilde{\mathbf{a}} = \mathbf{a} + \tilde{a} - a$$

(12)

$$\Delta F = \ln \mathrm{B}(a) - \ln \mathrm{B}(\tilde{a}) + \ln \mathrm{B}(\tilde{\mathbf{a}}) - \ln \mathrm{B}(\mathbf{a})$$
$$\Delta F := F[\tilde{P}(\theta) : P(\theta)] - F[P(\theta)]$$

This equation returns the difference in free energy we would have observed, had we started observing outcomes with simpler prior beliefs. This provides a criterion to accept or reject an alternative hypothesis – or reduced model structure – that is encoded by concentration parameters. For example, if a generative model contains likelihood matrices mapping from discrete (unknown) causes to (known) outcomes, one can parameterise the likelihood mapping with a categorical distribution and a Dirichlet prior. The posterior mapping from causes to outcomes then acquires a Dirichlet form that can be reduced during evidence accumulation. In other words, one can compare models with and without a mapping – between a particular cause and a particular outcome – by evaluating the change in free energy when the corresponding prior concentration parameter is set to zero. This enables a very efficient pruning of redundant parameters in models of discrete state states and outcomes; such as Markov decision processes and hidden Markov models. Intuitively, this form of structure learning enables one to simplify models by removing parameters that reduce the complexity to a greater degree than the implicit loss of accuracy (see Equation 4). We will see an example of this below, when used to model how the brain might implement this implicit form of structure learning.

In addition to Gaussian and Dirichlet distributions, Bayesian model reduction can be applied to a range of exponential family distributions. Table 1 outlines the form of reduced posterior and evidence for some common distributions. Note the similarity in form of the beta and Dirichlet expressions, and for multinomial and categorical distributions. This formal similarity is because these distributions are special cases of one another. Other special cases of these distributions, such as the binomial distribution (a special case of multinomial), have the same form. The categorical and multinomial reduced posteriors are expressed using a softmax function (σ) that exponentiates and then normalises its argument. A step-by-step derivation of the reduced evidence and posterior for a Gamma distribution is offered in the appendix, as an illustrative example.

Table 1 – Bayesian model reduction for common distributions

| Prior distribution | Sufficient statistics of reduced posterior | Log evidence ($\Delta F$) |
|---|---|---|
| **Beta** $$B(\alpha,\beta)^{-1}\theta^{\alpha-1}(1-\theta)^{\beta-1}$$ | $\tilde{\boldsymbol{\alpha}} = \boldsymbol{\alpha} + \tilde{\alpha} - \alpha$ <br> $\tilde{\boldsymbol{\beta}} = \boldsymbol{\beta} + \tilde{\beta} - \beta$ | $\ln B(\tilde{\alpha},\tilde{\beta}) - \ln B(\alpha,\beta)$ <br> $+ \ln B(\tilde{\boldsymbol{\alpha}},\tilde{\boldsymbol{\beta}}) - \ln B(\boldsymbol{\alpha},\boldsymbol{\beta})$ |
| **Categorical** $$e^{\theta \cdot \ln d} \quad s.t. \sum_i \theta_i = 1$$ | $\tilde{\mathbf{d}} = \sigma(\ln \tilde{d} + \ln \mathbf{d} - \ln d)$ | $\ln \tilde{d}_i - \ln d_i + \ln \mathbf{d}_i - \ln \tilde{\mathbf{d}}_i$ |
| **Dirichlet** $$B(a)^{-1} e^{(a-1)\cdot \ln \theta}$$ | $\tilde{\mathbf{a}} = \mathbf{a} + \tilde{a} - a$ | $\ln B(a) - \ln B(\tilde{a}) + \ln B(\tilde{\mathbf{a}}) - \ln B(\mathbf{a})$ |
| **Gamma** $$\Gamma(\alpha)^{-1}\beta^{\alpha}\theta^{\alpha-1}e^{-\beta\theta}$$ | $\tilde{\boldsymbol{\alpha}} = \boldsymbol{\alpha} + \tilde{\alpha} - \alpha$ <br> $\tilde{\boldsymbol{\beta}} = \boldsymbol{\beta} + \tilde{\beta} - \beta$ | $\boldsymbol{\alpha} \ln \boldsymbol{\beta} + \tilde{\alpha} \ln \tilde{\beta} - \alpha \ln \beta$ <br> $-(\boldsymbol{\alpha} + \tilde{\alpha} - \alpha)\ln(\tilde{\beta} + \boldsymbol{\beta} - \beta)$ <br> $+ \ln \Gamma(\alpha) - \ln \Gamma(\boldsymbol{\alpha}) - \ln \Gamma(\tilde{\alpha})$ <br> $+ \ln \Gamma(\boldsymbol{\alpha} + \tilde{\alpha} - \alpha)$ |
| **Gaussian** $$(2\pi\Sigma)^{-\frac{1}{2}} e^{-\frac{1}{2}\Sigma^{-1}(\eta-\theta)^2}$$ | $\tilde{C}^{-1} = \tilde{P} = P + \tilde{\Pi} - \Pi$ <br> $\tilde{\mu} = \tilde{C}(P\mu + \tilde{\Pi}\tilde{\eta} - \Pi\eta)$ | $\frac{1}{2}\ln\lvert \tilde{\Pi} P \tilde{C} \Sigma \rvert$ <br> $-\frac{1}{2}(\mu \cdot P\mu + \tilde{\eta}\cdot\tilde{\Pi}\tilde{\eta} - \eta\cdot\Pi\eta - \tilde{\mu}\cdot\tilde{P}\tilde{\mu})$ |

| Multinomial | | |
|---|---|---|
| $\dfrac{\Gamma(\sum_i \theta_i + 1)}{\prod_i \Gamma(\theta_i + 1)} e^{\theta \cdot \ln d} \quad s.t. \quad \sum_i \theta_i = n$ | $\tilde{\mathbf{d}} = \sigma(\ln \tilde{d} + \ln \mathbf{d} - \ln d)$ | $\ln \tilde{d}_i - \ln d_i + \ln \mathbf{d}_i - \ln \tilde{\mathbf{d}}_i$ |

# 6. Worked examples

In this section we provide worked examples of Bayesian model reduction in the context of static and dynamic, linear and non-linear models. Key Matlab routines used in the generation of the figures are given in brackets (e.g., **spm_dcm_bmr.m**). For full details of the model specifications for these examples, and to run the examples using Matlab, we refer the reader to the accompanying code documentation (see Software Note).

*6.1 Linear regression*

The first example demonstrates finding the optimal set of regressors (covariates or explanatory variables) needed to explain some set of observed data, in the context of a general linear model:

$$\hat{y} = X\theta + \varepsilon$$
$$\varepsilon \sim \mathcal{N}(0, \Sigma) \tag{13}$$
$$\Sigma = \exp(\lambda) \cdot I_N$$

Where vector $\hat{y} \in \mathbb{R}^N$ is the modelled data and the columns of design matrix $X \in \mathbb{R}^{N \times R}$ are the regressors with corresponding regression parameters $\theta \in \mathbb{R}^R$. Vector $\varepsilon \in \mathbb{R}^N$ is zero-mean additive noise with covariance matrix $\Sigma \in \mathbb{R}^{N \times N}$, controlled by a scalar hyperparameter $\lambda \in \mathbb{R}$, which is exponentiated to ensure positivity. To generate simulated data, we sampled a single observation from this model with 10 randomly generated (and orthogonalized) regressors in the design matrix with 10 corresponding randomly sampled parameters.

We then fitted a general linear model to the simulated data, using a variational Bayesian scheme (**spm_peb.m**) with Gaussian priors on the parameters and hyperparameters. This 'full' model had a design matrix that included the 10 regressors used to generate the data, plus an additional 10 randomly generated (orthogonal) regressors. We tested whether we could recover the original reduced model structure – i.e. the original 10 regressors – using Bayesian model reduction. We applied an automatic, iterative scheme (**spm_dcm_bmr_all.m**) to the full model, which evaluates reduced models with different combinations of the 20 parameters switched off. These parameters were switched off – i.e. fixed at values close to their prior expectation of zero – by replacing their original prior density, $\mathcal{N}(0,1)$ with a precise shrinkage prior, $\mathcal{N}(0,e^{-16})$.

Each reduced model's free energy was computed analytically using Bayesian model reduction (see Table 1), and the models with the greatest increase in free energy were retained for the next iteration of the (greedy) search. Figure 1A illustrates the model space from the final iteration of the search, in which eight parameters were variously combined to produce 256 reduced models with the greatest evidence. Parameters that were switched on are shown in white and parameters that were switched off are black. Figure 1B shows the relative free energy of each reduced model, compared to the full model. These results are seen more clearly in Figure 1C, which shows the posterior probability over models (assuming a uniform prior over models). Two models stand out – model 256 (84% probability), which had all eight of the candidate parameters switched on, and model 128 (14.6% probability), which had an additional parameter (number seven) switched off.

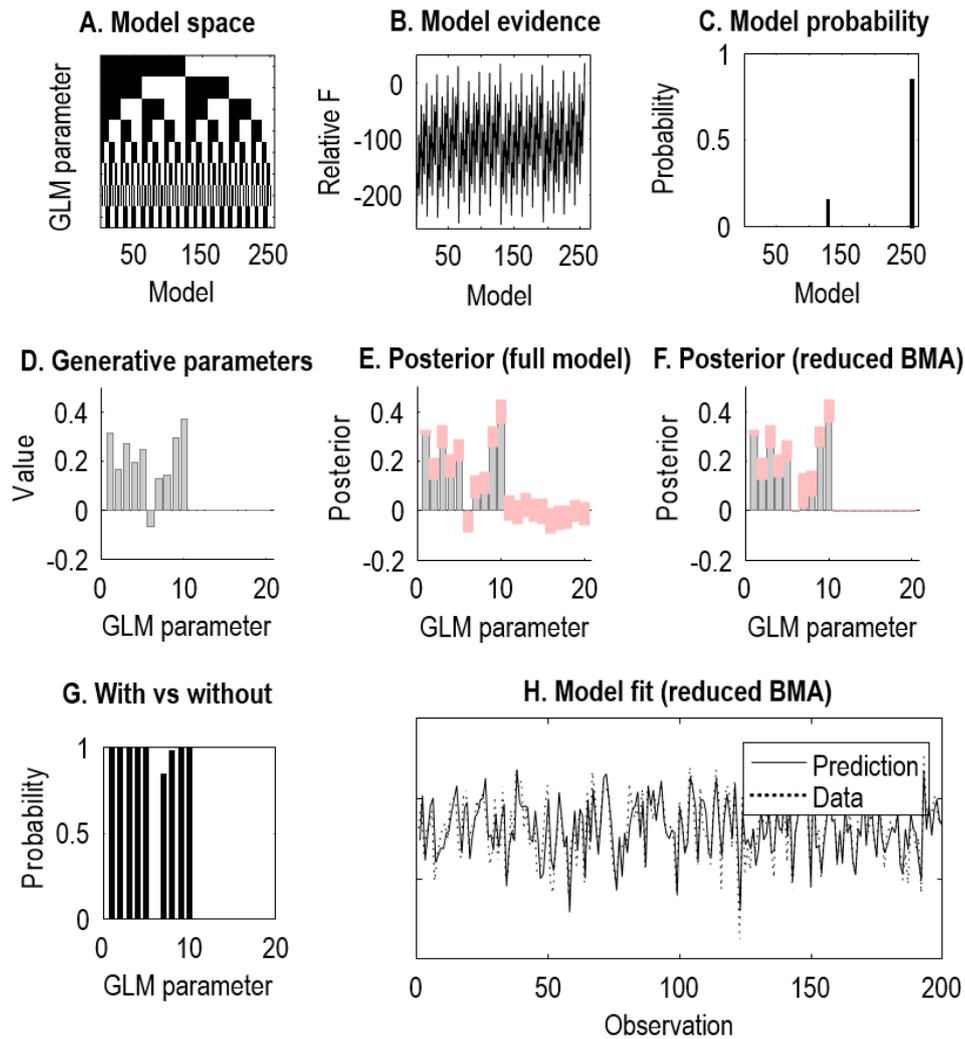

**Figure 1:** *Bayesian model reduction applied to simulated data from a general linear model.* A full model was defined with 20 regressors in its design matrix, and an iterative search was conducted over reduced models **A.** The set of 256 alternative models from the final iteration of the search. Each column is a model, each row a parameter. The colour denotes whether each parameter was on or off, i.e., white=on (uninformative prior) and black=off (fixed at a prior expectation of zero). **B**. The relative log model evidence, approximated by the free energy, relative to the full model, as evaluated using Bayesian model reduction. **C**. The posterior probability over models. The best model was number 256 (84% probability) and the second best was number 128 (14.6% probability). **D**. The parameters used to generate the simulated data. **E**. The posterior density over parameters from the full model. The height of the bars are the expected values, and pink error bars are 90% credible intervals. Covariance among parameters were also estimated but are not shown here. **F.** The Bayesian model average (BMA) of the parameters from all candidate models. Parameters 11-20 have been correctly switched off, as has parameter 6. **G**. Posterior probability for the presence of each parameter. Each bar shows the result of a comparison of all models which had a parameter switched on, versus all models with that parameter switched off. **H**. The prediction of the model (solid line) and residuals (dotted line) under the parameters shown in panel **F**.

The second row of Figure 1 compares the parameters used to generate the simulated data (Figure 1D) against the parameters recovered by the full model (Figure 1E) and the averaged parameters over the final 256 reduced models (Figure 1F). This was computed using Bayesian Model Averaging (BMA), meaning that the contribution of each model to the average was weighted by the model's posterior probability (Trujillo-Barreto et al., 2004). This form of averaging properly acknowledges uncertainty about the precise form or structure of the models entailed. From Figure 1F, it can be seen that that the redundant regression parameters (numbers 11-20) were correctly pruned from the model. Of the parameters that generated the data (numbers 1-10), one parameter was set to zero (number 6), indicating that its effect size was too small to be detected given the level of observation noise in the simulated data. In other words, including this parameter did not sufficiently increase the accuracy of the model to justify the additional complexity.

Figure 1G shows the probability of each parameter being present versus absent, which was computed by performing separate family-wise model comparisons for each parameter (Penny et al., 2010). In other words, for each parameter, the pooled evidence of all models that had the parameter switched on was compared against the pooled evidence of all models with the parameter switched off. This result demonstrates that we could be confident about the presence of all retained parameters with probability approaching unity, except for parameter seven which only had 85.2% probability. Finally, Figure 1H shows the actual data (dotted lines) against the data predicted by the model, under the parameters from the Bayesian model average.

In summary, this example illustrates the key procedures used in the context of Bayesian model reduction. A 'full' general linear model, with 20 parameters in play, was fitted to the data using a variational Bayes scheme, furnishing a posterior probability over parameters and the free energy approximation of the log evidence. Then, thousands of reduced models were scored (in a few seconds) using Bayesian model reduction with an automated iterative search. The posterior over the parameters from the best reduced models were summarised using Bayesian model averaging, and the probability for each parameter was computed using family-wise Bayesian model comparisons. Crucially, due to the use of Bayesian model reduction, it was only necessary to fit the full model to the data – the evidence and parameters for all other models were computed analytically, which takes around two seconds using a typical desktop computer.

*6.2 Gaussian mixture model*

The next example illustrates a Gaussian mixture model (GMM), which provides an opportunity to showcase a simple example of Bayesian model reduction applied to non-Gaussian distributions. The generative model takes the form shown in Figure 2 (assuming spherical unit covariance for simplicity). In brief, for each of $N$ data points, the model assumes a categorical variable ($s_n$) drawn from a categorical distribution (whose parameters are sampled from Dirichlet priors). This variable specifies the probability that the data point was generated by each cluster. Each value this variable can take is associated with a Gaussian density, with a different prior mean (initialised randomly). Solving this model involves finding the cluster responsibilities $Q(s_n)$ for each data point, the centres of each cluster $Q(\theta)$ and the proportion of data-points generated by each cluster $Q(D)$.

An important challenge in solving this kind of inference problem is deciding how many values the categorical variable can plausibly take (i.e., how many clusters are in play). To solve this, we start with more clusters than expected *a priori*, and eliminate clusters using Bayesian model reduction for the Dirichlet distribution over prior cluster probabilities $P(D)$. The key benefit of performing model reduction here is that we do not need to re-estimate the clusters with each cluster removed, before deciding whether to remove it. Rather, we can directly assess the reduced model evidence *as if* we had removed it. Figure 2 shows a sequence alternating between fitting a model with a set number of clusters and using Bayesian model reduction to evaluate the relative evidence for a set of models; each of which sets the prior probability of a given cluster to a small value. If the evidence is greater than that for the full model, the corresponding cluster is removed, until the correct number of clusters is recovered. This provides an example of how Bayesian model reduction can be used in the setting of categorical inference to select between alternative numbers of clusters.

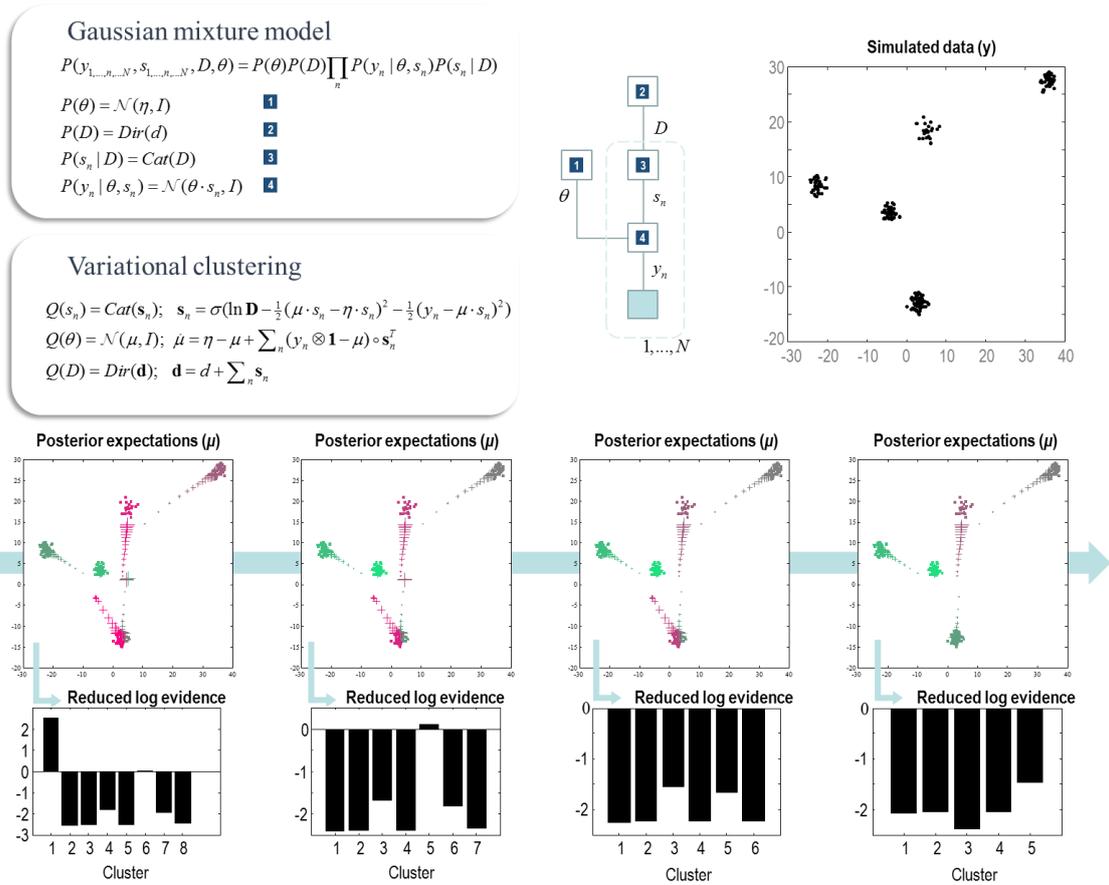

**Figure 2:** *Bayesian model reduction applied to simulated data from a Gaussian mixture model.* The **upper left panels** show the form of the Gaussian mixture model and variational clustering scheme used to assign each data point to its cluster. This is shown graphically as a factor graph in the **centre**. The upper right plot shows simulated data generated from this model, assuming five clusters. We then start with a model with eight clusters, with randomly generated prior expectations for their centres, and prune these away, until there is no further improvement in model evidence, as shown in the sequence of plots in the lower part of this figure. These plots show estimated values for $\mu$ as crosses (+) that increase in size for each gradient step. The colour of each datapoint indicates the most likely cluster that generated it. The reduced log evidence shown in the **lower row** corresponds to the log evidence of a model without each cluster, relative to a (full) model with all clusters in play. If this log evidence is greater than zero, the reduced model is accepted and the corresponding cluster is removed (i.e., the first cluster was removed on the first iteration and the fifth cluster was removed on the second iteration. The number of clusters decreases from six to five because we implemented an additional constraint that pairs of clusters whose likelihood distributions were less than three nats apart (as quantified with a KL-divergence) were merged. In this example, the correct number of clusters was recovered. Crucially, we do not need to fit any models with fewer clusters, as we can use model reduction to illustrate that there would be no further increase in model evidence had we done so. See the accompanying software code for an animated version of this figure.

*6.3 Dynamic causal modelling*

The previous two examples illustrated Bayesian model reduction (BMR) using static models; however, BMR can also be applied to the parameters of time series models based on differential equations. In neuroimaging, this is done routinely using Dynamic Causal Modelling (DCM); namely, the variational inversion of linear or nonlinear state space models of time series data (Daunizeau et al., 2011). In this setting, Bayesian model reduction enables the evidence for candidate network architectures to be scored rapidly; e.g., (Jafarian et al., 2019).

Here, we provide a simplified example. Consider a directed network or graph, comprising a set of interacting nodes (vertices) and edges between them. The nodes could be species in a food web, or companies on a stock market, or regions of the brain. Each node is equipped with a state $z_i$ quantifying, for example, the population size of a species, or the value of a company, or the post-synaptic firing rate of a neuronal population. Dynamic interactions play out on this network, giving rise to (noisy) measurements. The dynamics of this network may be modelled as follows:

$$\dot{z} = Az + Cu(t)$$
$$y = g(z) + \varepsilon$$

$$A_{ij} = \frac{\partial \dot{z}_i}{\partial z_j}$$

$$C_{mn} = \frac{\partial \dot{z}_m}{\partial u_n}$$

(14)

The first line of Eq 14 is a simple model (a Taylor approximation) of any dynamics of the form $\dot{z} = F_z(z, u(t))$, where $F_z$ is some unknown function. Variants of this model include the generalized Lotka-Volterra equations in ecology and economics (see Hofbauer and Sigmund, 1998), and the DCM for fMRI neural model used in neuroimaging (Friston et al., 2003). The adjacency or connectivity matrix $A \in \mathbb{R}^{N \times N}$ (the Jacobian, also called the community matrix) encodes the connections among the $N$ nodes, in which $A_{ij}$ is the influence from node *j* to node *i*. There are $R$ known perturbations that could drive activity in the network, encoded as time series in the columns of matrix $u(t)$ where $t$ indexes time, parameterized by matrix $C \in \mathbb{R}^{N \times R}$. The parameters $\theta = (A, C)$ are rate constants in units of Hz. The second line of Eq. 14 describes the generation of the measurements one would expect given the state vector $z$, via some linear or non-linear observation function $g$ and measurement error $\varepsilon$.

Here, we demonstrate the use of BMR to identify the adjacency matrix that best explains some data. We first simulated time series from an eight-node network (Figure 3). To generate the data, we configured matrix $A$ with excitatory connections in a feedforward direction (0.5Hz), reciprocal inhibitory connections in a feedback direction (-1Hz) and self-inhibition on each region to ensure stability (-0.25Hz). Activity in the network was driven by twenty 'events' of one second duration, with temporally jittered onsets that acted upon the first node; i.e., $C_{11} = 1\,\text{Hz}$. We added Gaussian observation noise sampled with a precision of 400, giving a signal-to-noise ratio of 8.18Db (based on the ratio of the summed squared signal to the sum squared noise). For simplicity, we set $g$ to the identity function, $g(z) = z$.

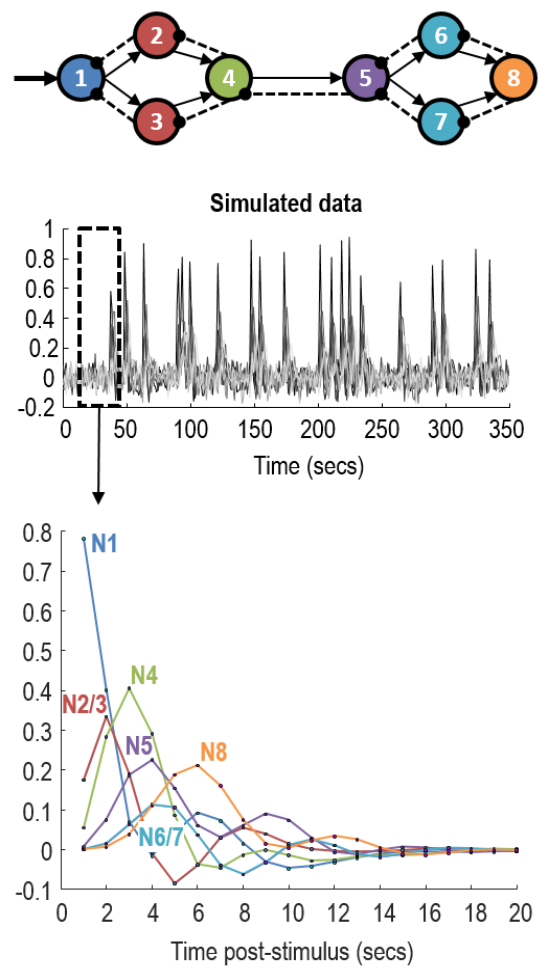

**Figure 3:** *Simulated data for Bayesian model reduction of a dynamical system.* Simulated data were generated from a simple differential equation model. **Top:** Configuration of eight nodes in the model used to generate the data. Solid arrows between nodes indicate positive (excitatory) values in connectivity matrix $A$ and dashed arrows indicate negative (inhibitory) values. The arrow entering node 1 indicates the driving input, encoded in matrix $C$ (see Equation 14). **Middle**: Simulated data generated by the model with the addition of observation noise. **Bottom**: Each node's state $z$ in response to a single input in the absence of noise. Note that temporal precedence is not the key data feature here – but rather the dynamic changes in amplitude, which resemble a damped oscillator due to the inhibitory self-connections on each node.

Next, we imagined that we did not know the connectivity structure that generated the data, and we wanted to infer it using the methods described above. To begin, we fitted a 'full' model to the data, with all possible connections among the 64 regions switched on. This corresponds to setting weakly informative Gaussian priors on the (connectivity) parameters of the adjacency matrix. A variational Laplace estimation scheme (**spm_nlsi_GN.m**) furnished the posterior probability density over the parameters of this full model and a free energy approximation of its evidence.

Next, we performed Bayesian model reduction. For illustrative purposes, we first compared a small set of hypothesis-driven reduced models and then performed an automated search over a large model space. There were three models in the hypothesis space (Figure 4, top). Model 1 was the full model, with all connections between all nodes informed by the data. Model 2 had certain connectivity parameters (elements of matrix $A$) fixed at zero, by replacing their default priors $\mathcal{N}(0,1/16)$ with precise shrinkage priors $\mathcal{N}(0,e^{-16})$. The structure of this model was the same as the one that generated the data. Model 3 had the same number of connections as model 2, but in a different configuration. We used Bayesian model reduction to compute the reduced free energy for each of the three candidate models. These are plotted in Figure 4 (bottom left) relative to the worst model. The bottom right plot is the posterior probability for each model (i.e., a softmax function over the free energies). As expected, model 2 was the winner, with posterior probability at ceiling. Thus, the original network structure was inferred after fitting only one model to the data (the full model) and analytically deriving the evidence for the two reduced models.

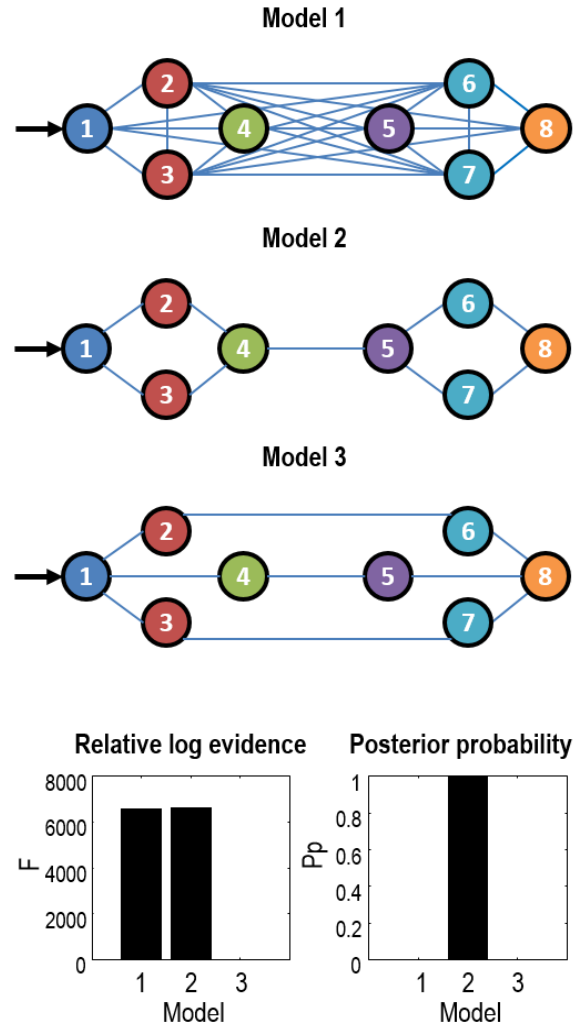

Figure 4: *Comparing pre-defined models using Bayesian model reduction.* After fitting a full model to the simulated data, three alternative models (which differed only in specification of their priors) were evaluated and compared. **Top panel:** the three candidate model architectures compared. Solid blue lines between nodes indicate connections that were allowed to vary, with prior $P(A_{ij}) = \mathcal{N}(0, 1/16)$, and the absence of a line between nodes indicates the connection was fixed at zero, with prior $P(A_{ij}) = \mathcal{N}(0, e^{-16})$. Model two matched the structure of the network used to generate the data, while model three had a different configuration of the same number of connections. **Bottom left:** The relative log model evidence (free energy) of each reduced model, computed using Bayesian model reduction. **Bottom right:** The corresponding posterior probability of each model.

Figure 5 shows an alternative application of Bayesian model reduction, using an automated search over reduced models. As with the linear regression example above, this involved iteratively pruning parameters from the full model, where doing so increased the free energy. The left panel of Figure 5 shows the parameters of the model used to generate the simulated data, and the middle panel shows the posterior density over the parameters of the full model, estimated using the variational Laplace scheme. The top row shows these parameters as bar charts, whereas the bottom row arranges the posterior expected values as adjacency matrices. It can be seen that the full model included many connectivity parameters that were not used to generate the data, with large error bars (90% credible intervals) overlapping zero. The right panel of Figure 5 shows the result of Bayesian model reduction. The original network architecture has been correctly recovered, with the exception of the connection from node 7 to node 8, which was pruned from the model. This illustrates the fact that Bayesian model comparison will always favour the simplest model that can explain the data. Here, node 8's activity could be sufficiently explained by the input it received from node 6, without warranting the additional complexity cost of an input from node 7.

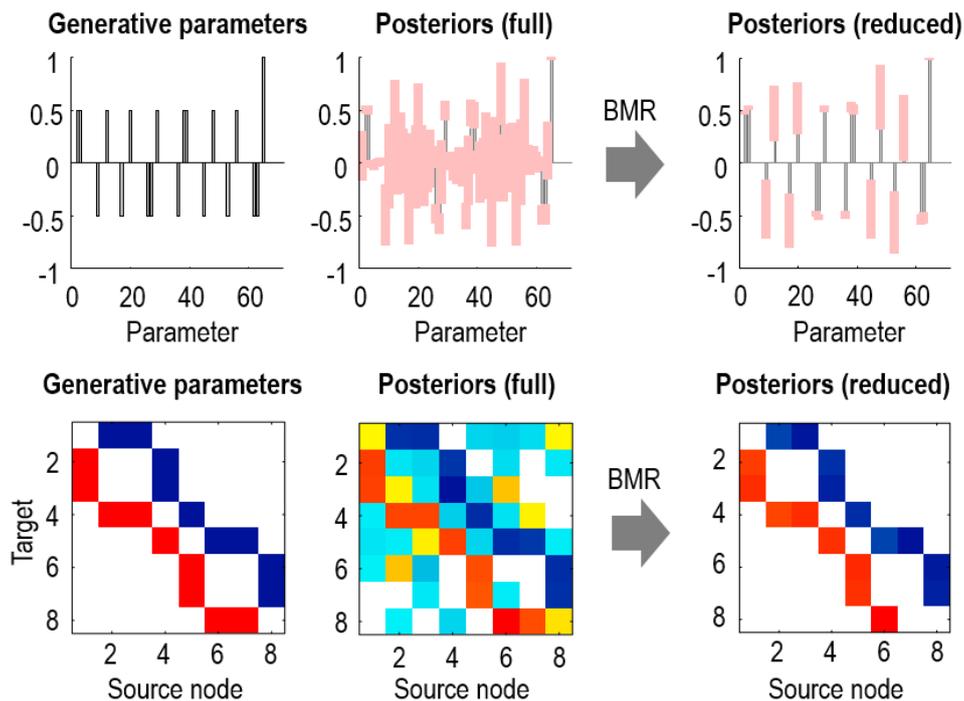

**Figure 5**: *Automatic search over reduced dynamical models using Bayesian model reduction*. The top row illustrates the model parameters as bars (with 90% credible intervals where available), the bottom row shows the expected values as adjacency matrices. **Left column**: parameters used to generate the simulated data. These are ordered into outgoing connections from node 1, then node 2 *etc*. The final parameter is from matrix *C* (the external input driving the system). **Middle column**: Estimated connectivity from the full model, which had all connections among the eight nodes switched on. **Right column**: Bayesian model average over the 256 best models from an automatic (greedy) search. The recovered parameters closely match those used to generate the data, with the exception of the connection from region 7 to region 8, which was not needed to explain the data.

## 7. Empirical examples

Bayesian model reduction has proved useful in recent years in the fields of computational neuroscience and biology. We close by briefly reviewing some of the different types of models to which Bayesian model reduction has been applied. Inevitably, most of these examples are from our own lab and those of our collaborators, but throughout this paper we hope to emphasise the generality of the approach.

*7.1 Bayesian model comparison and structure learning*

As illustrated above, a useful application of Bayesian model reduction is to score very large model spaces, after inverting a single parent model under relatively uninformative priors. This was capitalised upon in Friston et al. (2011), in which the (dynamic causal) model detailed in Section 6.3 was extended to model the dynamics of the brain at rest (Figure 6A). This required the use of differential equations to model network dynamics, and a biophysically detailed observation model to describe the translation from hidden states (neuronal responses) to functional magnetic resonance imaging (fMRI) time series data. When inferring the connectivity structure of a large network such as this, the number of connections and their combinations can clearly become prohibitively large; thereby calling for an efficient model or structure learning scheme. In this setting, by using Bayesian model reduction, one can evaluate thousands of candidate models in a few seconds. This example used Gaussian posteriors over continuous states. For a detailed survey of recent work using Bayesian model reduction for structure learning of dynamic models, please see (Jafarian et al., 2019).

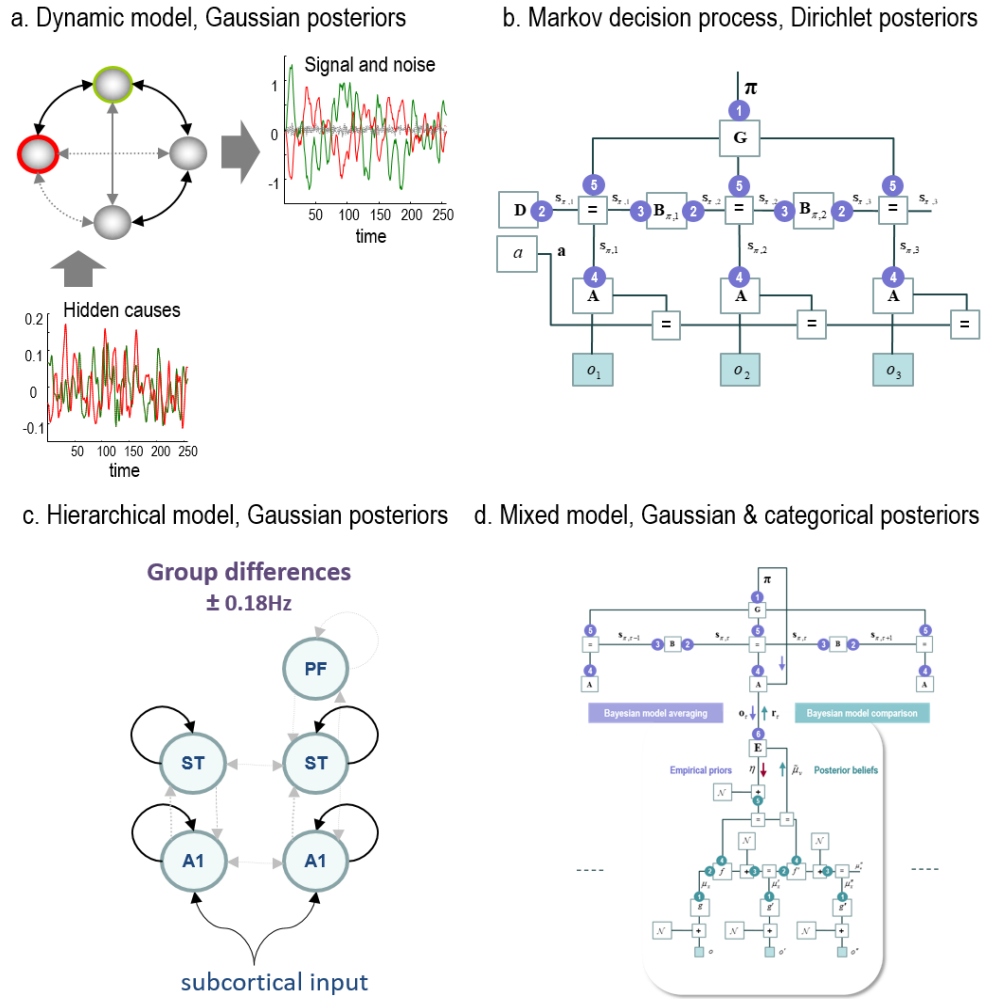

Figure 6 *Example models used with Bayesian model reduction in neurobiology and neuroimaging.* **Panel a**. Continuous state space neural network model (top left) used to simulate functional magnetic resonance imaging (fMRI) data (top right of panel a). Stochastic model inversion (generalized filtering) was used to perform the difficult dual problem of estimating connectivity parameters (thin black and grey arrows) and unobserved causes of neural activity (bottom left of panel a). Bayesian model reduction was then used to infer the connectivity architecture that generated the data (black arrows). Adapted from (Friston et al., 2011). **Panel b**. Graphical representation of a generative model used to simulate rule learning. This Forney factor graph shows the conditional dependencies implied by the generative model. The variables in white circles constitute (hyper) priors, while the blue circles contain random variables. This format shows how outcomes (o) are generated from hidden states (s) that evolve according to probabilistic transitions (B), which depend on policies (π). The probability of a particular policy being selected depends upon its expected free energy (G). Bayesian model reduction was used on matrix A, which encodes the likelihood mapping from hidden states to outcomes. Adapted from (Friston et al., 2017a). **Panel c**. Continuous state space model of the brain's auditory pathway, used to simulate electromagnetic responses from multiple subjects, in the context of between-group differences. The hierarchical model consisted of continuous state space models for each subject at the first level and a general linear model at the second level. The objective was to infer commonalities and between-group differences in particular connectivity parameters. Adapted from (Friston et al., 2016). **Panel d**. This figure uses Forney factor graphs to illustrate a message passing scheme that integrates discrete (Markov decision process) and continuous (state space) models. The upper half of this graph corresponds to a Markov decision process – using the same parameterisation has the generative model in panel b. The continuous model in the lower half was a state space model whose parameters were determined by the discrete outcome states of the Markov decision process. The communication between the lower (continuous) and higher (discrete) levels of the model used Bayesian model reduction. Please see (Friston et al., 2017b, Friston et al., 2017c) for details.

*7.2 Bayesian model reduction in biology*

Bayesian model reduction has not only been useful for data analysis and model comparison, it has also been leveraged in the context of computational neuroscience and theoretical neurobiology. Both the elimination of redundant parameters inherent in Bayesian model comparison – and the optimization of deep, hierarchical models have figured in this context. A nice example of a biological process eliminating redundant parameters is sleep; via the elimination or regression of synaptic connections in the brain (Tononi and Cirelli, 2006).

This basic notion has been modelled using synthetic (*in silico*) subjects who model contingencies in their world (Friston et al., 2017a). Using Bayesian model reduction, it is fairly easy to show that simply 'thinking about things' off-line can minimise model complexity and lead to much more efficient learning. In this example, a curious agent was simulated that spontaneously learnt rules governing the sensory consequences of its actions. On each trial of a virtual experiment, the agent was shown three coloured circles and could act by directing its gaze. When ready, the agent responded with a choice of colour and was given feedback on whether their response was correct or incorrect. To succeed at this task, the agent had to infer the rules determining the feedback it received (e.g., if the circle at the top was red, the correct response was the colour on the left).

Crucially, synthetic subjects were able to enhance their abductive inference (i.e., reasoning), when they engaged Bayesian model reduction after each exposure to sensory data (i.e., feedback). Here, the model space was equipped with extra (hyperprior) constraints that defined the class of models the simulated subjects entertained (the class of likelihood mappings between discrete hidden states and observable outcomes). By minimising the complexity of the ensuing models – using Bayesian model reduction – they proved more generalisable to new data and experience; thereby increasing the efficiency of inference and learning. The generative model used to simulate rule learning was a discrete time (partially observed) Markov Decision Process (Figure 6b) with Dirichlet priors and posteriors over the parameters of the categorical distributions.

Thus far, we have focused on the use of reduced free energy for model comparison, averaging and selection. The next example illustrates using the same technology to finesse problems in hierarchical, deep or empirical Bayesian models (Kass and Steffey, 1989).

*7.3 Hierarchical or deep modelling*

Bayesian model reduction has proven effective in the inversion of deep or hierarchical models. For example, say one had inverted some highly nonlinear, high-dimensional state space model of several subjects and now wanted to make inferences about model parameters at the between subject level; e.g., (Friston et al., 2015). This would necessarily entail some form of hierarchical modelling; however, it would be nice not to have to re-invert each subject-specific model every time some between-subject parameter changed. In this hierarchical setting, the reduced free energy functional above finds a particularly powerful application, because it summarises everything that one needs to know at any level of the hierarchical model, in terms of optimizing the approximate posteriors of all levels above. In short, it converts a full hierarchical inversion problem into a succession of Bayesian model reduction problems – in which the posterior beliefs at successively higher levels of the model are optimized based upon the priors and posteriors of the level below.

More formally, consider a hierarchical model expressed in terms of conditional distributions over a succession of unknown model parameters:

$$P(y,\theta) = P(y|\theta_1)P(\theta_1|\theta_2)P(\theta_2|\theta_3)\ldots P(\theta_n) \quad (15)$$

Imagine now that we inverted the first level of the model, while ignoring any higher constraints from supraordinate levels. We then inverted the first two levels of the model, ignoring third and higher levels – and so on.

$$F_1[P(\theta_1)] = E_{Q_1}[\ln P(y|\theta_1)] - D_{KL}[Q(\theta_1) \| P(\theta_1)]$$
$$F_2[P(\theta_2)] = E_{Q_1}[\ln P(y|\theta_1)] - D_{KL}[Q(\theta_1) \| E_{Q_2}[P(\theta_1|\theta_2)]] - D_{KL}[Q(\theta_2) \| P(\theta_2)]$$
$$F_3[P(\theta_3)] = \ldots$$

$$\Rightarrow \quad (16)$$

$$F_1[P(\theta_1)] = E_{Q_1}[\ln P(y|\theta_1)] - D_{KL}[Q(\theta_1) \| P(\theta_1)]$$
$$F_2[P(\theta_2)] = F_1[E_{Q_2}[P(\theta_1|\theta_2)]] - D_{KL}[Q(\theta_2) \| P(\theta_2)]$$
$$F_3[P(\theta_3)] = \ldots$$

One can see from the above expressions that the free energy of the full hierarchical model can be expressed recursively in terms of the reduced free energy at all subordinate levels. In other words, as we add hierarchical constraints – with each extra level of the model – the corresponding free energy can be evaluated in terms of the extra complexity incurred at the new level and the reduced free energy of the previous level under the empirical prior afforded by the new level:

$$F_i[P(\theta_i)] = \underbrace{F_{i-1}[E_{Q_i}[P(\theta_{i-1} | \theta_i)]]}_{\text{Accuracy}} - \underbrace{D_{KL}[Q(\theta_i) \| P(\theta_i)]}_{\text{Complexity}}$$

$$\begin{aligned} F_0 &:= \ln P(y | \theta_1) \\ F_1 &:= F[P(\theta_1)] \\ F_{i+1}[P(\theta_{i+1})] &:= F[E_{Q_i}[P(\theta_i | \theta_{i+1})] : P(\theta_i)] \end{aligned} \quad (17)$$

In this setting, the reduced free energy functional plays the role of an augmented likelihood that has all the necessary information to evaluate the accuracy at any level of the model. This means that we only need to optimize the accuracy (using the free energy functional of posteriors and priors from the lower level) and the complexity due to the parameters at the current level in question.

Heuristically, the use of the reduced free energy functional enables one to replace the sort of analysis used in mixed effects modelling of within and between-subject effects with a summary statistic approach, in a strictly feedforward fashion; i.e., passing sufficient statistics upwards from a lower-level to the next. Here, the reduced free energy functional summarises the evidence for beliefs about parameters at the higher level in terms of the posterior over the parameters of the lower level and the empirical priors from the higher level.

Under Laplace (i.e., Gaussian) or Dirichlet assumptions about the form of the posterior, nothing really changes mathematically when adopting this summary statistic approach. However, the computation times can be greatly improved. This is because one does not have to revisit all lower levels of the hierarchical model to update the posterior beliefs under successive empirical prior afforded by higher levels. This was illustrated in (Friston et al., 2016), in which a neural network model (Figure 6C) with 158 neuronal parameters was fitted to simulated electroencephalogram (EEG) data, for each of 16 simulated subjects. In half the subjects, certain neuronal parameters were altered to introduce a between-group difference (as one might see in a comparison of patients and healthy controls). The parameters of interest were then taken to the second (between-subject) level, where a general linear model encoded commonalities and between-group differences. Specific connections expressing between-group differences were correctly identified using this Parametric Empirical Bayes (PEB) scheme. Crucially, the estimation of posterior beliefs over parameters at the between subject level – and ensuing Bayesian model reduction – only took a few seconds; despite the fact that the inversion of each subject's dynamic causal model could take a minute or so.

*7.4 Deep (graphical) models in neurobiology*

A second example from neurobiology speaks to the (variational) message passing between levels of a deep or hierarchical model. Perhaps the best example of this in the literature describes a synthetic subject that can perform a simple form of (iconographic) reading. Crucially, the hierarchical model here involved a separation of temporal scales and a mixture of continuous (lower level) and discrete (higher-level) states spaces (Figure 5D). The communication between the lower (continuous) and higher (discrete) levels of the model used Bayesian model reduction, so that the log evidence for a particular hidden state in the (Markov decision process) model was provided by the free energy lower bound using the reduced free energy. In this case, the model had Gaussian posteriors over Gaussian states. These were used to provide the evidence for a set of models, each of which was associated with a categorical state at a higher level of the model. This is similar to the Gaussian mixture model outlined above. The evidence for each categorical state was combined with their (categorical) priors to compute categorical posterior beliefs in the standard way. Please see (Friston et al., 2017b, Friston et al., 2017c) for details.

# 8. Conclusion

In summary, we have reviewed Bayesian model reduction and the free energy functional upon which it rests. This special case of Bayesian model comparison rests upon specified analytic forms for the prior and posterior densities of any given model. In the setting of approximate Bayesian inference, this usually mandates a variational approach to model fitting, and subsequent comparison. Within this setting – and under the constraint that all interesting models can be specified in terms of prior constraints on a full or parent model – Bayesian model reduction has proven extremely useful. In particular, it enables one to rapidly score the evidence for large model spaces, for example to compare models with different sparsity structures (i.e., with different mixtures of parameters 'switched off' by fixing them at their prior expectations). We have highlighted an application of Bayesian model reduction that is proving particularly useful in neuroimaging – the inversion of deep or hierarchical models, where the reduced free energy provides an efficient summary of the evidence needed to inform beliefs about parameters in successively higher levels. Another potential application, not pursued here, would be to investigate the sensitivity of parameters to their priors, by plotting their posteriors as a function of reduced priors; thereby conducting a sensitivity or robustness analysis (Giordano et al., 2016). Some key areas of application and utility have been highlighted in the hope that the basic ideas could be adopted elsewhere.

## 9. Appendix

This appendix offers a step-by-step derivation of the expression for the reduced evidence and posterior for a Gamma distribution. The priors and posteriors for a Gamma distribution can be expressed:

$$\Gamma(\alpha,\beta) = \Gamma(\alpha)^{-1}\beta^{\alpha}\theta^{\alpha-1}e^{-\beta\theta}$$

$$\begin{aligned}P(\theta) &= \Gamma(\alpha,\beta) \\ Q(\theta) &= \Gamma(\boldsymbol{\alpha},\boldsymbol{\beta}) \\ \tilde{P}(\theta) &= \Gamma(\tilde{\alpha},\tilde{\beta})\end{aligned} \quad (18)$$

Substituting these into Eq. 9 gives, for the reduced free energy:

$$\begin{aligned}\Delta F &= \ln E_Q\left[\frac{\tilde{P}(\theta)}{P(\theta)}\right] \\ &= \ln \frac{\boldsymbol{\beta}^{\boldsymbol{\alpha}}\tilde{\beta}^{\tilde{\alpha}}\Gamma(\alpha)}{\beta^{\alpha}\Gamma(\boldsymbol{\alpha})\Gamma(\tilde{\alpha})}\int \theta^{\boldsymbol{\alpha}+\tilde{\alpha}-\alpha-1}e^{(\beta-\boldsymbol{\beta}-\tilde{\beta})\theta}d\theta \\ &= \ln \frac{\boldsymbol{\beta}^{\boldsymbol{\alpha}}\tilde{\beta}^{\tilde{\alpha}}\Gamma(\alpha)\Gamma(\boldsymbol{\alpha}+\tilde{\alpha}-\alpha)}{\beta^{\alpha}\Gamma(\boldsymbol{\alpha})\Gamma(\tilde{\alpha})(\beta-\boldsymbol{\beta}-\tilde{\beta})^{\boldsymbol{\alpha}+\tilde{\alpha}-\alpha}} \\ &= \boldsymbol{\alpha}\ln\boldsymbol{\beta} + \tilde{\alpha}\ln\tilde{\beta} - \alpha\ln\beta - (\boldsymbol{\alpha}+\tilde{\alpha}-\alpha)\ln(\beta-\boldsymbol{\beta}-\tilde{\beta}) \\ &\quad + \ln\Gamma(\alpha) + \ln\Gamma(\boldsymbol{\alpha}+\tilde{\alpha}-\alpha) - \ln\Gamma(\boldsymbol{\alpha}) - \ln\Gamma(\tilde{\alpha})\end{aligned} \quad (19)$$

The third equality here uses the fact that the term inside the integral is proportional to a Gamma distribution, so the integral is the associated normalising constant. For the reduced posterior:

$$\ln \tilde{Q}(\theta) = \ln Q(\theta) - \ln P(\theta) + \ln \tilde{P}(\theta) - \ln E_Q \left[ \frac{\tilde{P}(\theta)}{P(\theta)} \right]$$

$$\begin{aligned}
&\tilde{\boldsymbol{\alpha}} \ln \tilde{\boldsymbol{\beta}} + (\tilde{\boldsymbol{\alpha}} - 1) \ln \theta - \tilde{\boldsymbol{\beta}} \theta - \ln \Gamma(\tilde{\boldsymbol{\alpha}}) \\
&= \boldsymbol{\alpha} \ln \boldsymbol{\beta} + (\boldsymbol{\alpha} - 1) \ln \theta - \boldsymbol{\beta} \theta - \ln \Gamma(\boldsymbol{\alpha}) \\
&\quad + \tilde{\alpha} \ln \tilde{\beta} + (\tilde{\alpha} - 1) \ln \theta - \tilde{\beta} \theta - \ln \Gamma(\tilde{\alpha}) \\
&\quad - \alpha \ln \beta - (\alpha - 1) \ln \theta + \beta \theta + \ln \Gamma(\alpha) \\
&\quad - \Delta F \\
\\
&= (\boldsymbol{\alpha} + \tilde{\alpha} - \alpha - 1) \ln \theta - (\boldsymbol{\beta} - \tilde{\beta} + \beta) \theta - (\boldsymbol{\alpha} + \tilde{\alpha} - \alpha) \ln(\tilde{\beta} + \boldsymbol{\beta} - \beta) - \ln \Gamma(\boldsymbol{\alpha} + \tilde{\alpha} - \alpha)
\end{aligned} \qquad (20)$$

$$\tilde{\boldsymbol{\alpha}} = \boldsymbol{\alpha} + \tilde{\alpha} - \alpha$$
$$\tilde{\boldsymbol{\beta}} = \boldsymbol{\beta} - \tilde{\beta} + \beta$$

## 10. Software note

The figures in this note can be reproduced using routines are available as Matlab code in the SPM academic software: http://www.fil.ion.ucl.ac.uk/spm/. The simulations in this paper can be reproduced (and customised) via a graphical user interface: by typing >> DEM and selecting the appropriate (e.g., **rule learning**) demo. Table 2 lists the functions in the SPM software implementing Bayesian model reduction.

Table 2 – Matlab functions in SPM for Bayesian model reduction

| Function name | Priors / posteriors |
| --- | --- |
| spm_gamma_log_evidence | Gamma |
| spm_log_evidence | Gaussian |
| spm_MDP_log_evidence | Dirichlet, Beta |
| spm_multinomial_log_evidence | Multinomial, Binomial, Bernoulli, Categorical |

## 11. Acknowledgements

The Wellcome Centre for Human Neuroimaging is supported by core funding from Wellcome [203147/Z/16/Z]. KJF is funded by Wellcome (Ref: 088130/Z/09/Z). TP is funded by a Rosetrees Ph.D. Fellowship. Thank you to Michael Hopkins for helpful feedback on the manuscript.

## 12. Disclosure statement

The authors have no disclosures or conflict of interest.

## 13. References

BEAL, M. J. 2003. *Variational algorithms for approximate Bayesian inference*, PhD thesis, University College London.

COLLINS, A. G. & FRANK, M. J. 2013. Cognitive control over learning: creating, clustering, and generalizing task-set structure. *Psychol Rev,* 120**,** 190-229.

DAUNIZEAU, J., DAVID, O. & STEPHAN, K. E. 2011. Dynamic causal modelling: a critical review of the biophysical and statistical foundations. *Neuroimage,* 58**,** 312-322.

DAUWELS, J. On variational message passing on factor graphs.  Information Theory, 2007. ISIT 2007. IEEE International Symposium on, 2007. IEEE, 2546-2550.

FOX, C. W. & ROBERTS, S. J. 2012. A tutorial on variational Bayesian inference. *Artificial intelligence review,* 38**,** 85-95.

FRISTON, K., MATTOUT, J., TRUJILLO-BARRETO, N., ASHBURNER, J. & PENNY, W. 2007. Variational free energy and the Laplace approximation. *Neuroimage,* 34**,** 220-234.

FRISTON, K. & PENNY, W. 2011. Post hoc Bayesian model selection. *Neuroimage,* 56**,** 2089-99.

FRISTON, K., ZEIDMAN, P. & LITVAK, V. 2015. Empirical Bayes for DCM: A Group Inversion Scheme. *Front Syst Neurosci,* 9**,** 164.

FRISTON, K. J., HARRISON, L. & PENNY, W. 2003. Dynamic causal modelling. *Neuroimage,* 19**,** 1273-302.

FRISTON, K. J., LI, B. J., DAUNIZEAU, J. & STEPHAN, K. E. 2011. Network discovery with DCM. *Neuroimage,* 56**,** 1202-1221.

FRISTON, K. J., LIN, M., FRITH, C. D., PEZZULO, G., HOBSON, J. A. & ONDOBAKA, S. 2017a. Active Inference, Curiosity and Insight. *Neural Comput,* 29**,** 2633-2683.

FRISTON, K. J., LITVAK, V., OSWAL, A., RAZI, A., STEPHAN, K. E., VAN WIJK, B. C. M., ZIEGLER, G. & ZEIDMAN, P. 2016. Bayesian model reduction and empirical Bayes for group (DCM) studies. *Neuroimage,* 128**,** 413-431.

FRISTON, K. J., PARR, T. & DE VRIES, B. 2017b. The graphical brain: belief propagation and active inference. *Network Neuroscience,* 1**,** 381-414.

FRISTON, K. J., ROSCH, R., PARR, T., PRICE, C. & BOWMAN, H. 2017c. Deep temporal models and active inference. *Neuroscience & Biobehavioral Reviews,* 77**,** 388-402.

GIORDANO, R., BRODERICK, T., MEAGER, R., HUGGINS, J. & JORDAN, M. 2016. Fast robustness quantification with variational Bayes. *arXiv preprint arXiv:1606.07153*.

HOFBAUER, J. & SIGMUND, K. 1998. *Evolutionary games and population dynamics*, Cambridge university press.

JAFARIAN, A., ZEIDMAN, P., LITVAK, V. & FRISTON, K. 2019. Structure Learning in Coupled Dynamical Systems and Dynamic Causal Modelling. *arXiv preprint arXiv:1904.03093*.

KASS, R. E. & RAFTERY, A. E. 1995. Bayes factors. *Journal of the american statistical association,* 90**,** 773-795.

KASS, R. E. & STEFFEY, D. 1989. Approximate Bayesian inference in conditionally independent hierarchical models (parametric empirical Bayes models). *Journal of the American Statistical Association,* 84**,** 717-726.

KSCHISCHANG, F. R., FREY, B. J. & LOELIGER, H.-A. 2001. Factor graphs and the sum-product algorithm. *IEEE Transactions on information theory,* 47**,** 498-519.

LOELIGER, H.-A. 2002. Least squares and Kalman filtering on Forney graphs. *Codes, Graphs, and Systems.* Springer.

MACKAY, D. J. 1995. Free energy minimisation algorithm for decoding and cryptanalysis. *Electronics Letters,* 31**,** 446-447.

MITTER, S. K. & NEWTON, N. J. 2003. A variational approach to nonlinear estimation. *SIAM journal on control and optimization,* 42**,** 1813-1833.

PENNY, W. D., STEPHAN, K. E., DAUNIZEAU, J., ROSA, M. J., FRISTON, K. J., SCHOFIELD, T. M. & LEFF, A. P. 2010. Comparing families of dynamic causal models. *PLoS computational biology,* 6**,** e1000709.

RANGANATH, R., GERRISH, S. & BLEI, D. Black box variational inference.  Artificial Intelligence and Statistics, 2014. 814-822.

RANGANATH, R., TRAN, D. & BLEI, D. Hierarchical variational models.  International Conference on Machine Learning, 2016. 324-333.

REZENDE, D. J. & MOHAMED, S. 2015. Variational inference with normalizing flows. *arXiv preprint arXiv:1505.05770*.

ROWEIS, S. & GHAHRAMANI, Z. 1999. A unifying review of linear Gaussian models. *Neural computation,* 11**,** 305-345.

SALAKHUTDINOV, R., TENENBAUM, J. B. & TORRALBA, A. 2013. Learning with hierarchical-deep models. *IEEE Trans Pattern Anal Mach Intell,* 35**,** 1958-71.

SALIMANS, T., KINGMA, D. & WELLING, M. Markov chain monte carlo and variational inference: Bridging the gap.  International Conference on Machine Learning, 2015. 1218-1226.

SAVAGE, L. J. 1972. *The foundations of statistics*, Courier Corporation.

SCHMIDHUBER, J. Curious model-building control systems.  Neural Networks, 1991. 1991 IEEE International Joint Conference on, 1991. IEEE, 1458-1463.

SCHMIDHUBER, J. 2010. Formal theory of creativity, fun, and intrinsic motivation (1990–2010). *IEEE Transactions on Autonomous Mental Development,* 2**,** 230-247.

TENENBAUM, J. B., KEMP, C., GRIFFITHS, T. L. & GOODMAN, N. D. 2011. How to grow a mind: statistics, structure, and abstraction. *Science,* 331**,** 1279-85.

TERVO, D. G. R., TENENBAUM, J. B. & GERSHMAN, S. J. 2016. Toward the neural implementation of structure learning. *Curr Opin Neurobiol,* 37**,** 99-105.

TONONI, G. & CIRELLI, C. 2006. Sleep function and synaptic homeostasis. *Sleep medicine reviews,* 10**,** 49-62.

TRUJILLO-BARRETO, N. J., AUBERT-VAZQUEZ, E. & VALDES-SOSA, P. A. 2004. Bayesian model averaging in EEG/MEG imaging. *Neuroimage,* 21**,** 1300-19.

VERDINELLI, I. & WASSERMAN, L. 1995. Computing Bayes factors using a generalization of the Savage-Dickey density ratio. *Journal of the American Statistical Association,* 90**,** 614-618.

WOOLRICH, M. W., JBABDI, S., PATENAUDE, B., CHAPPELL, M., MAKNI, S., BEHRENS, T., BECKMANN, C., JENKINSON, M. & SMITH, S. M. 2009. Bayesian analysis of neuroimaging data in FSL. *Neuroimage,* 45**,** S173-86.

YEDIDIA, J. S., FREEMAN, W. T. & WEISS, Y. 2005. Constructing free-energy approximations and generalized belief propagation algorithms. *IEEE Transactions on information theory,* 51**,** 2282-2312.

ZORZI, M., TESTOLIN, A. & STOIANOV, I. P. 2013. Modeling language and cognition with deep unsupervised learning: a tutorial overview. *Front Psychol,* 4**,** 515.